
\magnification=\magstep1
\let\bigbold=\bf
\let\Bbb=\cal

\newcount\EQNO      \EQNO=0
\newcount\FIGNO     \FIGNO=0
\newcount\REFNO     \REFNO=0
\newcount\SECNO     \SECNO=0
\newcount\SUBSECNO  \SUBSECNO=0
\newcount\FOOTNO    \FOOTNO=0
\newbox\FIGBOX      \setbox\FIGBOX=\vbox{}
\newbox\REFBOX      \setbox\REFBOX=\vbox{}
\newbox\RefBoxOne   \setbox\RefBoxOne=\vbox{}

\expandafter\ifx\csname normal\endcsname\relax\def\normal{\null}\fi

\def\Eqno{\global\advance\EQNO by 1 \eqno(\the\EQNO)%
    \gdef\label##1{\xdef##1{\nobreak(\the\EQNO)}}}
\def\Fig#1{\global\advance\FIGNO by 1 Figure~\the\FIGNO%
    \global\setbox\FIGBOX=\vbox{\unvcopy\FIGBOX
      \narrower\smallskip\item{\bf Figure \the\FIGNO~~}#1}}
\def\Ref#1{\global\advance\REFNO by 1 \nobreak[\the\REFNO]%
    \global\setbox\REFBOX=\vbox{\unvcopy\REFBOX\normal
      \smallskip\item{\the\REFNO .~}#1}%
    \gdef\label##1{\xdef##1{\nobreak[\the\REFNO]}}}
\def\Section#1{\SUBSECNO=0\advance\SECNO by 1
    \bigskip\leftline{\bf \the\SECNO .\ #1}\nobreak}
\def\Subsection#1{\advance\SUBSECNO by 1
    \medskip\leftline{\bf \ifcase\SUBSECNO\or
    a\or b\or c\or d\or e\or f\or g\or h\or i\or j\or k\or l\or m\or n\fi
    )\ #1}\nobreak}
\def\Footnote#1{\global\advance\FOOTNO by 1 
    \footnote{\nobreak$\>\!{}^{\the\FOOTNO}\>\!$}{#1}}
\def\SameFootnote{$\>\!{}^{\the\FOOTNO}\>\!$}

\def\References{\bigskip\centerline{\bf REFERENCES}
                \smallskip\copy\REFBOX}
\def\NewRefPage{\setbox\RefBoxOne=\vbox{\unvcopy\REFBOX}%
		\setbox\REFBOX=\vbox{}%
		\def\References{\bigskip\centerline{\bf REFERENCES}
                		\nobreak\smallskip\nobreak\copy\RefBoxOne
				\vfill\eject
				\smallskip\copy\REFBOX}%
		\def\NewRefPage{}}


\def\MultiRef#1{\global\advance\REFNO by 1 \nobreak\the\REFNO%
    \global\setbox\REFBOX=\vbox{\unvcopy\REFBOX\normal
      \smallskip\item{\the\REFNO .~}#1}%
    \gdef\label##1{\xdef##1{\nobreak[\the\REFNO]}}}

\def\today{\number\day\space\ifcase\month\or
  January\or February\or March\or April\or May\or June\or
  July\or August\or September\or October\or November\or December\fi
  \space\number\year}
\rightline{ADP 95-43/M36 \hfill 23 August 1995 }
\rightline{gr-qc/9601034 \hfill (revised 13 January 1996)}

\def\SR{{\Bbb S}^2 \times {\Bbb R}^2}
\def\rmax{r_{\rm max}}

\def\om{\omega}
\def\DelT{\Delta t}

\bigskip


\null\bigskip
\centerline{\bigbold THE ROTATING QUANTUM VACUUM}
\bigskip

\centerline{Paul C. W. Davies}
\centerline{\it Dept.\ of Physics \& Mathematical Physics,
		University of Adelaide, Adelaide, SA 5005, AUSTRALIA}
\centerline{\tt pdavies{\rm @}physics.adelaide.edu.au}
\medskip
\centerline{Tevian Dray
\Footnote{Permanent address is Oregon State University.}
}
\centerline{\it Dept.\ of Physics \& Mathematical Physics,
		University of Adelaide, Adelaide, SA 5005, AUSTRALIA}
\centerline{\it Department of Mathematics, Oregon State University,
		Corvallis, OR  97331, USA}
\centerline{\tt tevian{\rm @}math.orst.edu}
\medskip
\centerline{Corinne A. Manogue
\SameFootnote
}
\centerline{\it Dept.\ of Physics \& Mathematical Physics,
		University of Adelaide, Adelaide, SA 5005, AUSTRALIA}
\centerline{\it Department of Physics, Oregon State University,
		Corvallis, OR  97331, USA}
\centerline{\tt corinne{\rm @}physics.orst.edu}

\bigskip\bigskip
\centerline{\bf ABSTRACT}
\midinsert
\narrower\narrower\noindent
We derive conditions for rotating particle detectors to respond in a variety
of bounded spacetimes and compare the results with the folklore that particle
detectors do not respond in the vacuum state appropriate to their motion.
Applications involving possible violations of the second law of thermodynamics
are briefly addressed.
\endinsert
\bigskip

\centerline{PACS: 04.62.+v}

\Section{INTRODUCTION}

The nature of the quantum vacuum continues to be the subject of surprising
discoveries.  One of the more curious properties to be discussed in recent
years is the prediction that an observer who accelerates in the conventional
quantum vacuum of Minkowski space will perceive a bath of radiation, while an
inertial observer of course perceives nothing.  In the case of linear
acceleration, for which there exists an extensive literature, the response of
a model particle detector mimics the effect of its being immersed in a bath of
thermal radiation (the so-called Unruh effect).  The investigation of rigid
{\it rotation}, as opposed to linear acceleration, seems however to have been
somewhat limited
[\MultiRef{JD Pfautsch, PhD Dissertation, The University of Texas at Austin,
(1981); 
\hfill\break
JD Pfautsch, Phys.\ Rev.\ {\bf D24} (1981) 1491;
\hfill\break
JR Letaw \& JD Pfautsch, J. Math.\ Phys.\ {\bf 23} (1982) 425.}%
\label\Pfautsch,%
\MultiRef{PG Grove \& AC Ottewill, J. Phys.\ {\bf A16} (1983) 3905.}%
\label\Grovedetect,%
\MultiRef{T Padmanabhan, Astrophysics and Space Sciences {\bf 83} (1982) 247.}%
\label\Paddy,%
\MultiRef{CA Manogue, Phys.\ Rev.\ {\bf D35} (1987) 3783.}]%
, in spite of the fact that experimental tests of the theory
are more feasible for the rotating case
\Ref{JS Bell \& JM Leinaas, Nucl.\ Phys.\ {\bf B212} (1983) 131;
\hfill\break
JS Bell \& JM Leinaas, Nucl.\ Phys.\ {\bf B284} (1987) 488.
}%
.%

The problem of rotation is a very deep one in physics which goes back at least
to Newton.  Through his famous bucket experiment, Newton sought to demonstrate
that rotation is an absolute effect taking place in a substantival space,
later to be identified with the aether.  By contrast, Mach argued that
rotation is purely relative to the distant matter in the universe.  Then with
the inception of the theory of relativity the aether was abandoned, but the
question of whether rotation is relative or absolute was not settled by the
theory of relativity, and the status of Mach's principle remains contentious
even today.

In recent years there has been an attempt to revive a type of aether concept
by appealing to the quantum vacuum
[\MultiRef{BS DeWitt in
{\bf General Relativity:  An Einstein Centenary Survey},
ed. SW Hawking \& W Israel (Cambridge University Press, Cambridge, 1979)
680.}\label\DeWitt,%
\MultiRef{
DW Sciama, P Candelas, \& D Deutsch, Adv.\ Phys.\ {\bf 30} (1981) 327.}]%
, and even to link this quantum vacuum with Machian effects
\Ref{B Haisch, A Rueda, \& HE Puthoff, Phys.\ Rev.\ {\bf A49} (1994) 678.}%
.  We might envisage the quantum vacuum playing the role of Newton's
substantival space.  It is therefore of some interest to investigate the
effects of rotation in quantum field theory.

In what follows we attempt to progress this discussion by investigating the
properties of particle detectors moving in circular paths.  The results turn
out to involve some remarkable and subtle issues, especially with regard to
the definition of a rotating quantum vacuum state.

The paper is organized as follows.  In Section 2, we first review previous
results about rotating particle detectors in unbounded Minkowski space.  We
then add a cylindrical boundary in Section 3, and consider compact spaces in
Section 4.  The possibility of violating the second law of thermodynamics is
discussed in Section 5, and in Section 6 we conclude by discussing the
physical interpretation of our results.  We use units with $\hbar = c = 1$,
our metric signature convention is $(+ - - -)$, and we treat massless scalar
fields for simplicity.

\Section{UNBOUNDED MINKOWSKI SPACE}

We first summarize the results of Letaw and Pfautsch \Pfautsch.  Consider a
DeWitt model particle detector \DeWitt\ moving in a circular path of radius
$r$ at constant angular velocity $\Omega$ in the conventional Minkowski
(inertial) vacuum state.  Here, and in all the cases we shall be investigating
later, the response of the detector is independent of time.  According to the
standard theory, the probability of excitation of a detector per unit
(detector) proper time is given by
$$
	{{\cal F}(E)\over T} = \int\limits_{-\infty}^\infty d\Delta\tau \,
		e^{-iE\Delta\tau} G^+ \left( x(\tau),x(\tau') \right)
\Eqno$$\label\Response
where $G^+$ is the positive frequency Wightman function.  The Minkowski space
Wightman function in cylindrical polar coordinates $x=(t,r,\theta,z)$ is
$$
	G^+(x,x') = {1\over4\pi^2} \,\, {1\over
			(t-t'-i\epsilon)^2
			-(r^2+r'^2-2rr'\cos(\theta-\theta')+(z-z')^2) }
~~.
\Eqno$$\label\GPolar
The detector's trajectory is given by $r={\rm constant}$, $z={\rm constant}$
and $\theta=\Omega t$, so that $G^+$ is a function only of the difference,
$\Delta\tau$, in proper time.  The response function is then given by the
Fourier transform integral
$$
	{{\cal F}(\bar{E})\over T}
	= {\sqrt{1-r^2\Omega^2}\over4\pi^2}
	  \int\limits_{-\infty}^\infty d\DelT \,\,
	  {e^{-i\bar{E}\DelT}\over
		(\DelT - i\epsilon)^2 - 4r^2\sin^2({\Omega\DelT\over2})}
\Eqno$$\label\FPolar
where $\bar{E}=E\sqrt{1-r^2\Omega^2}$ corresponds to the Lorentz re-scaling
$\DelT = t-t' = \Delta\tau/\sqrt{1-r^2\Omega^2}$ arising from the
transformation between proper time $\tau$ and coordinate time $t$.  This
expression was evaluated numerically by Pfautsch.  The fact that it is nonzero
is already an important result, analogous to the famous linear acceleration
case, where the detected spectrum of particles is thermal.  Although
Pfautsch's spectrum is reminiscent of a Planck spectrum, the similarity with
the thermal linear case is only superficial.
	
The Minkowski vacuum state is defined with respect to the usual field modes
based on mode solutions $u_{qmk}$ of the wave equation in inertial
coordinates.
$$
u_{qmk}={1\over 2\pi\sqrt{2\omega}}\, \, 
                J_{m} \left( qr \right)
	        \, e^{im\theta}\, e^{ikz} \, e^{-i\om t}
\Eqno$$
where $\omega^2=q^2+k^2$.

The transformation from Minkowski to rotating coordinates appears at first
sight to have only a trivial implication for the modes of the quantum field:
We may readily transform to a rotating coordinate system, by letting
$\hat\theta = \theta - \Omega t$.  Mode solutions $\hat{u}_{qmk}$ of the
rotating wave equation are then found to be identical to the conventional
(non-rotating) Minkowski modes, and can be obtained, up to normalization, by
the replacement $\hat\omega = \omega - m\Omega$.  In particular, the Bogolubov
transformation between the rotating and non-rotating modes is trivial ---
amounting to a simple relabeling of modes.  Furthermore, since the two sets of
modes are identical, they have identical norms, so that no mixing of positive
and negative ``frequencies'' occurs irrespective of a possible change of sign
between $\omega$ and $\hat\omega$.  It is the norms, not the frequencies, of
the rotating modes which determine the commutation relations of the associated
creation and annihilation operators.

If these rotating modes are used to define a rotating vacuum state, it
coincides with the conventional Minkowski vacuum.  This is in contrast to the
case for linear acceleration 
\Ref{SA Fulling, Phys.\ Rev.\ {\bf D7} (1973) 2850.}%
, where the so-called Rindler
(Fulling) vacuum contains Minkowski particles and vice versa.  Furthermore,
there is a general expectation that a family of model particle detectors
``adapted'' to the coordinate system on which the quantized modes are based
will reveal the particle content of those modes.  Thus, in the Rindler case, a
set of uniformly accelerating particle detectors sharing common asymptotes
will give zero response in the Rindler vacuum state, and will give a
consistent thermal response to the Minkowski vacuum state \DeWitt.  We might
therefore expect a set of rotating detectors to similarly reveal the state of
a rotating quantum field.  This is not the case here: Since the inertial and
rotating vacuum states are identical, the response of a rotating particle
detector to the co-rotating vacuum will be the same as \FPolar\ above, and we
have the curious result that a particle detector adapted to the rotating
vacuum apparently does not behave as if it is coupled to a vacuum state at
all.

Before accepting this odd result, however, we must confront a serious problem.
For a given angular velocity $\Omega$, there will be a maximum radius
$\rmax=1/\Omega$ (the light cylinder) beyond which a point at fixed $r$ and
$\theta$ will be moving faster than light, i.e.\ the rotating Killing vector
$\partial_t-\Omega\partial_\theta$ becomes spacelike.  It is far from clear
what to make of the rotating field modes, or even the notion of Bogolubov
transformation, outside this limiting light circle.  This complication
compromises any straightforward attempt to construct rotating quantum states
in terms of the mode solutions just described.

To circumvent this problem we eliminate the light cylinder altogether.  We
accomplish this in two ways: first, in Section 3, by introducing a cylindrical
boundary within the light cylinder and confining the quantum field to the
bounded region, and then, in Section 4, by making the space compact.

\Section{ROTATING DETECTOR IN A BOUNDED DOMAIN}

The Wightman function \GPolar\ inside a cylinder of radius $a < \rmax$, on
which the field satisfies vanishing (i.e.\ Dirichlet) boundary conditions, can
be written as a discrete mode sum
$$
	G^+(x,x') =
		\sum_{m=-\infty}^\infty \, \sum_{n=1}^\infty
		\, \int\limits_{-\infty}^\infty dk \, {N^2\over\om} 
		\, J_{m} \left( {\xi_{mn}r\over a} \right)
		\, J_{m} \left( {\xi_{mn}r'\over a} \right)
	        \, e^{im(\theta-\theta')}
		\, e^{ik(z-z')} \, e^{-i\om(t-t')}
\Eqno$$\label\GCyl
where
$$N = {1 \over 2\pi a \, |J_{m+1}(\xi_{mn})|} ~~,$$
$\xi_{mn}$ is the $n$th zero of the Bessel function $J_{m}(x)$, and
$$
	\om = \sqrt{{\xi_{mn}^2\over a^2} + k^2}
~~.
\Eqno$$
As is appropriate for the positive frequency Wightman function, we assume
$\omega>0$.  For a particle detector moving in a circular path of radius $r$
about the axis of the cylinder with uniform angular velocity $\Omega$, we set
$r={\rm constant}$, $z={\rm constant}$ and $\theta=\Omega t$.  Substituting
from \GCyl\ into \Response, we obtain for the detector response per unit time
$$
{{\cal F}(E)\over T}=
	\sqrt{1-r^2\Omega^2}
	\sum_{m=-\infty}^\infty \, \sum_{n=1}^\infty
	\, \int\limits_{-\infty}^\infty dk \, {N^2\over\om} 
	\, J_{m}^2 \left( {\xi_{mn}r\over a} \right)
	\, \int\limits_{-\infty}^\infty d\DelT \,
	\, e^{-i( \bar{E} + \om - m\Omega) \DelT}
~~.
\Eqno$$\label\CylResponse
We may immediately perform the $\DelT$ integral to obtain the delta
function
$$
	2\pi \, \delta \left( \bar{E}+ \om - m\Omega \right)
	\equiv 2\pi \, \delta \left( 
		\bar{E}+\sqrt{{\xi_{mn}^2\over a^2} + k^2}-m\Omega \right)
~~.
\Eqno$$\label\CylDelta
Since $\bar{E}>0$ and $\Omega>0$ by assumption, the argument of the delta
function will vanish only if $m>0$.  Furthermore, since the radial modes are
now discrete, for each $m$ there will be a lowest value of $\om$, namely
$\xi_{m1}/a$ ($k$ is a continuous momentum variable in the $z$ direction and
can be zero). Therefore, the detector will fail to respond unless
$$
	m\Omega a > \xi_{m1}
~~.
\Eqno$$
But we know from the properties of Bessel functions that the zeros satisfy the
inequality
$$
	\xi_{mn} > m
\Eqno$$
which leads to the important conclusion that unless $\Omega a > 1$ the
rotating detector will remain inert.  (The same conclusion would hold even if
we had chosen Neumann boundary conditions for the field.)

Physically, our result may be interpreted as follows.  Imagine a point on the
boundary at $r=a$, co-rotating with the detector; if the point does not exceed
the speed of light then the particle detector remains unexcited, i.e.\ it
registers a vacuum.  Now we may repeat the above calculation for the case of a
vacuum state that co-rotates with the detector.  Because the Bogolubov
transformation between the rotating and non-rotating vacuum states is trivial,
the result of the calculation will differ only by the formal relabeling
$\hat\omega=\omega-m\Omega$.  Thus we may conclude that, subject to the above
speed of light restriction, a rotating particle detector co-rotating with a
rotating vacuum state registers the absence of quanta.  This is in accord with
our intuition.  But the speed of light restriction is precisely what is needed
to render the concept of a rotating vacuum unabiguous.  If the boundary lies
outside the crucial light circle, then the ambiguity returns.

The spectrum of radiation that will be detected by a rotating detector above
the critical rotation rate is given by inserting the delta function \CylDelta\
into \CylResponse, and performing the $k$ integral
$$
{{\cal F}(E)\over T} =
	2\pi \sqrt{1-r^2\Omega^2} 
	\sum_{m} \, \sum_{n}
	\, N^2 \, {J_{m}^2 ({\xi_{mn}r\over a})\over
		\sqrt{(m\Omega-\bar{E})^2-{\xi_{mn}^2\over a^2}}}
\Eqno$$\label\CylSpectrum
where the sums in $m$ and $n$ run over those values in the range
$\left[ 0,\infty\right)$ for which ${\xi_{mn}\over a} < m\Omega - \bar{E}$. 
We have not bothered to evaluate \CylSpectrum\ numerically, but we expect on
general grounds for the spectrum to have a sawtooth form at low energies
\Ref{LH Ford, Phys.\ Rev.\ {\bf D38} (1988) 528.}%
.%


\Section{$\SR$ AND THE EINSTEIN UNIVERSE}

In Section 3, we removed the ambiguity associated with faster-than-light
rotation by confining the quantum field within a boundary.  We can also
consider the case of a compact space, wherein the field is automatically
confined.  To illustrate this possibility, we discuss two examples.  For the
first example, we consider the detector to be rotating in a plane at uniform
angular velocity $\Omega$ in the static space with topology $\SR$ and metric
$$
	ds^2 = dt^2 -
		a^2 \left( d\theta^2 + \sin^2\theta \, d\phi^2 \right)-dz^2
\Eqno$$
where $a$ is the (constant) radius of the 2-sphere.  The massless scalar wave
equation has solutions
$$
  u_{lmk} = {N\over\sqrt\omega} Y_{lm}(\theta,\phi) \, e^{ikz} e^{-i\omega t}
\Eqno$$
where $N$ is an unimportant normalization factor.  The eigenfrequencies are
$$
	\omega = \sqrt{{l(l+1)+2\xi \over a^2} + k^2}
\Eqno$$
where $\xi$ is the curvature coupling, so that the corresponding delta
function in place of \CylDelta\ is
$$
	\delta\left( \bar{E} +
		\sqrt{{l(l+1)+2\xi \over a^2} +k^2} - m\Omega \right)
\Eqno$$
where $k$ is a continuous variable that can be zero and $l$,$m$ are integers
satisfying $l\ge0$, $|m|\le l$.  Again it is clear that the detector will
remain inert unless $m>0$.  The lowest value of $\omega$ for which excitation
will occur is given by $l=1$, $m=1$.  The argument of the delta function will
then vanish only if
$$
     a\Omega > \cases{\sqrt{2}&($\xi=0$)\cr \sqrt{7\over3}&($\xi={1\over6}$)}
~~.
\Eqno$$\label\Bands

This can be translated into the condition that a rotating detector will remain
unexcited if a co-rotating point situated at the ``equator'' (relative to the
origin of coordinates around which the detector circulates) moves at less than
the speed of light.  For the case of minimal ($\xi=0$) or conformal
($\xi=1/6$) coupling, excitation occurs only if the said point moves at
$\sqrt{2}$ or $\sqrt{7/3}$ times the speed of light respectively.  Since the
detector itself must move at less than the speed of light, we see that there
is a band around the equator for which no uniformly rotating detector
responds.  For minimal coupling, this band consists of all latitudes less than
$\pi/4$.  This includes geodesic detectors at the equator, which are not
accelerating, some of which respond for unphysical couplings $\xi<-1/2$.

In our second example, the detector rotates in the Einstein static universe
with metric
$$\eqalign{
  ds^2
  &= dt^2 - a^2 \left( d\chi^2 + \sin^2\chi \, (
     d\theta^2 + \sin^2\theta \, d\phi^2 ) \right)\cr
  }$$
where $a$ is the (constant) radius of the universe.  It is unnecessary to deal
with the specific form of the field modes
\Ref{LH Ford, Phys.\ Rev.\ {\bf D14} (1976) 3304.}
, as the method used in the previous examples is quite general, and works in
any stationary axisymmetric spacetime.  The only important information is that
the eigenfrequencies are
$$
\omega={\sqrt{(n+1)^2+6\xi-1 \over a^2}}
\Eqno$$
where $n,m$ label the eigenmodes, with $n \ge 0$ and $m \le n$.

Repeating the analogous steps leading up to \CylDelta, we obtain on performing
the time integration
$$
	2\pi \, \delta
	  \left( \bar{E} + {\sqrt{(n+1)^2+6\xi-1 \over a^2}} - m\Omega \right)
~~.
\Eqno$$
The lowest value of $\omega$ for which excitation will occur is given by
$n=1$, $m=1$.  We may immediately conclude that the detector will remain
unexcited unless
$$
     a\Omega > \cases{\sqrt{3}&($\xi=0$)\cr ~~2&($\xi={1\over6}$)}
\Eqno$$\label\Einstein
(cf.\ \Bands) thus yielding the same type of band structure as for $\SR$.

Interestingly, for the case $\xi=1/6$ the Wightman function for the Einstein
universe has been given in closed form
[\MultiRef{R Critchley, PhD Thesis, University of Manchester (1976)},%
\MultiRef{ND Birrell \& PCW Davies, {\bf Quantum fields in curved space},
(Cambridge University Press, Cambridge, 1982) 123.}]%
, so that we may write down an alternative expression for the detector's
response in terms of a Fourier transform
$$
{{\cal F}(E)\over T} =
	-{1\over16\pi^2} \int\limits_{-\infty}^\infty d\DelT
	\, {e^{-i\bar{E}\DelT} \over
	\sin^2\left({\DelT-i\epsilon\over 2a}\right)
	- {r^2\over a^2} \sin^2\left({\Omega \DelT\over2}\right)}
~~.
\Eqno$$
The integral can be performed by residues, with a contour along the real axis
closed by a semicircle in the lower half $\DelT$-plane.  Because of the
presence of the $i\epsilon$ factor in the denominator, the poles that lie
along the real axis will not contribute.  This includes a pole near $\DelT=0$.
However, there can be additional poles on the imaginary axis (and hence a
nonzero detector response) if there is a second solution (other than
$\DelT=0$) to the equation
$$
	\sinh\left({\DelT\over 2a}\right) =
		\pm {r\over a} \sinh\left({\Omega \DelT \over 2}\right)
~~.
\Eqno$$\label\Sinh
Noting that $r/a<1$, and writing $\Omega \DelT = \Omega a\, (\DelT/a)$, we
see that for such a solution to exist requires $\Omega a > 1$, which agrees
with our above result apart from a puzzling factor of $2$.

It is also of interest to note that when $r=a$ the (conformally-coupled)
detector is moving along a geodesic on the equator of the 3-sphere with
respect to our coordinate system.  Under these circumstances there can be no
solution other than $t=0$ to \Sinh.  Hence, the detector will never respond,
regardless of its velocity (which must be slower than the speed of light).


\Section{NEGATIVE ENERGY \& THE SECOND LAW OF THERMODYNAMICS}

Many cases are known where the energy density of a quantum vacuum state is
negative in some spacetime region, for instance, the Casimir vacuum state
between parallel conducting plates, the Rindler vacuum within the Rindler
wedge, the space outside a static star in the Boulware vacuum, and the vacuum
region outside a straight cosmic string.  If an observer moves through such a
region, the possibility arises of scooping up enough negative energy to
violate the second law of thermodynamics, or black hole cosmic censorship.
For example, if the negative energy is allowed to accumulate within an opaque
box containing a gas of excited atoms at a finite temperature, the negative
energy may cool the gas by de-exciting the atoms.

Investigations suggest that, in the case of linear motion, either the
symmetries of the problem imply that the flux of negative energy is zero in
the frame of the moving observer, or the spatial extent of the negative energy
region is so circumscribed that the total accumulated negative energy is
limited by the Ford bound
\Ref{LH Ford, Phys.\ Rev.\ {\bf D48} (1993) 776.}\label\Ford
, and can do no harm.  Thus in the Casimir case, motion parallel to the plates
corresponds to zero energy flux, while motion perpendicular to the plates will
not be possible for long because the observer will collide with one of the
plates, thereby saving the second law.

However, if we allow for circular motion, the situation appears to be
different; the observer may travel through a negative energy region ad
infinitum.  Ford has considered the example of geodesic observers orbiting on
circular paths around Schwarschild black holes in the Unruh vacuum state
\Ford.  Between the radii $r=3M$ and $r\approx 5M$ the energy flux in the
frame of the observer is always negative.  On the face of it, such scenarios
seem to threaten the second law.  However, if a rotating observer ``sees''
radiation, then this ``rotation radiation'' may excite the contents of a hot
box by more than the negative energy flux de-excites them, saving the second
law.  If this is the correct resolution of the problem for unbounded
spacetimes, then the fact that the excitation of a rotating detector can be
suppressed by bounding the space is deeply disturbing.  For example, would a
hot box slowly rotating around a cosmic string lying along the axis of a
reflecting cylinder violate the second law?  The following heuristic argument
suggests yes.  The presence of a cylinder of large radius would be unlikely to
have much effect on the energy density close to the string, which is large and
negative.  As there is no bound on how slowly the second law may be violated,
the hot box can always rotate around the string slowly enough for the cylinder
to suppress excitation.

\Section{PHYSICAL INTERPRETATION \& OPEN QUESTIONS}

The principle result of our investigation is that, whenever ambiguities in the
appropriate definition of a ``rotating vacuum'' state associated with the
region beyond the light cylinder are circumvented, the detector fails to
respond.  However, when the angular velocity of the detector is above a
certain threshold, excitation occurs.  The threshold corresponds, within a
factor of order unity, to the situation that a co-rotating imaginary point in
the spacetime exceeds the speed of light.

How can one interpret this response physically, in the case that excitation
does occur?  One is tempted to reason as follows.  In the rotating frame, an
observer would perceive a (albeit nonthermal) bath of radiation, and the
excitation of the detector corresponds to a quantum of this bath ``being
detected''.  In the inertial frame no such bath exists.  Instead, the detector
emits a quantum into the field, making an upward transition as a result.
However, because of the bounded spatial domain, the lowest field mode has a
finite energy, so that the detector will respond only when the detector is
able to excite the lowest mode.  If the angular velocity is too low, the
detector is unable to overcome this threshold and emit a quantum of sufficient
energy to excite the field.  It therefore remains inert.  Unfortunately, this
simple interpretation is flawed.  Consider the example of a detector moving in
a circular path between the plates of a Casimir system, in a plane parallel to
the plates.  The time integral leads in this case to a delta function of the
form
$$
2\pi \, \delta\left( \bar{E} + \sqrt{q^2 + \left( \pi n\over a\right)^2} 
- m\Omega \right)
\Eqno$$
where $a$ is the plate separation. The lowest energy mode corresponds to
$q=0$, $n=1$ and is nonzero due to the discretization of the modes in the $z$
direction.  However, because there is no upper bound on $m$, the argument of
the delta function can be zero for arbitrarily small values of the angular
velocity $\Omega$.  This example shows that the threshold effect controlling
when the detector responds is not simply related to the existence of a
discrete field state of lowest energy.  

Therefore we are left with a number of unanswered questions.  When a moving
detector responds, where does the energy of excitation ``come from'', and what
effect will the transition have on the energy of the field?  Will the field
have more or less energy after the excitation, and will any change in field
energy be confined to within the light cylinder?  There is considerable
confusion in the literature about the energetics of detector response even in
the linearly accelerating case
[\MultiRef{WG Unruh \& RM Wald, Phys.\ Rev.\ {\bf D29} (1984) 1047.},%
\MultiRef{S Takagi, Prog. Theor. Phys. {\bf 72} (1984) 505.},%
\MultiRef{T Padmanabhan, Class.\ Quant.\ Grav.\ {\bf 2} (1985) 117.},%
\MultiRef{PG Grove, Class.\ Quant.\ Grav.\ {\bf 3} (1986) 801.}]%
, i.e.\ whether the quantum field gains or loses energy, in what region of
spacetime the quantum field energy changes, and whether the accelerating
agency supplies the energy needed to excite the detector.

The rotating case is different.  When the co-rotating vacuum is unambiguous,
nothing happens, and the considerations above presumably do not apply.  The
detector responds precisely when questions arise as to what state should be
defined as the co-rotating vacuum, making it even more difficult to provide a
physical interpretation.  It has been previously proposed \Grovedetect\ to
distinguish between {\it particle emission} (Rindler case) and {\it radiation
backreaction} (rotating case) when giving the inertial observer's explanation
of the detector response.  As this distinction is itself ultimately based on
constructing the Bogolubov coefficients between the rotating and inertial
vacuum states, it is not yet clear whether this argument can be applied.

It is interesting to speculate whether it might be possible to define some
other state for which there is a space-filling family of rotating detectors
which do not respond.  Here, we note two facts which may have some bearing on
this question.  If the usual positive frequency Wightman function with
$\omega>0$ is replaced by one with $\hat\omega>0$, then, even in unbounded
Minkowski space, the argument of the delta function in \CylDelta\ is
$\bar{E}+\hat\omega$, which is always positive.  This appears to describe a
situation in which a co-rotating detector would not respond.  However, the
usual correspondence between Wightman functions and vacuum states fails here
as the ``state'' corresponding to this Wightman function is not in the usual
Fock space.  One could instead consider differentially rotating detectors,
which are thus in non-stationary motion.  But then the observers would not
move along integral curves of a single Killing vector field as required by the
usual quantization procedures.  If such a state were to exist, and a suitable
quantization procedure defined [\MultiRef{A Ashtekar
\& A Magnon,
Proc.\ Roy.\ Soc. {\bf A346} (1975) 375.}%
,\MultiRef{Tevian Dray, Ravi Kulkarni, \& Corinne A. Manogue, 
Gen.\ Rel.\ Grav.\ {\bf 24} (1992) 1255.}]%
, it would have a strong claim to be designated as
the rotating vacuum.

In some respects the light cylinder resembles the boundary of the Rindler
system.  The Rindler coordinate system is well defined within the spacetime
wedge delineated by the null rays $x=\pm t$.  Outside this wedge the Rindler
spatial coordinates become timelike, i.e. Rindler observers would be moving
faster than light.  Similarly, in the case of a rotating black hole, it is
well known that geodesic observers in the region known as the ergosphere,
close to the hole, are spun around faster than light relative to unaccelerated
observers at infinity.  The outer boundary of the ergosphere --- the so-called
static limit surface --- is reminiscent of the light cylinder for the rotating
vacuum.  In this case, we would expect a particle detector moving along a
(non-rotating) Killing trajectory to respond to particles created via the
Unruh-Starobinskii effect
[\MultiRef{AA Starobinskii, Sov.\ Phys.\ JETP {\bf 37} (1973) 28.},%
\MultiRef{WG Unruh, Phys.\ Rev.\ {\bf D10} (1974) 3194.}]%
.  However, cases have been found, for instance a relativistic star
\Ref{AL Matacz, PCW Davies, AC Ottewill,
Phys.\ Rev.\ {\bf D47} (1993) 1557.}%
, for which there is an ergosphere but no horizon, and for which the
Unruh-Starobinskii effect is absent.  We hope to report on the response of
co-moving detectors for some of these examples in a future publication.

\bigskip\leftline{\bf ACKNOWLEDGMENTS}\nobreak

TD and CAM would like to thank the Department of Physics and Mathematical
Physics at the University of Adelaide, and especially the newly formed
Green-Hurst Institute for Theoretical Physics, for kind hospitality during
their sabbatical visit.  This work was partially supported by NSF Grant
PHY-9208494 (CAM \& TD) and a Fulbright Grant (TD) under the auspices of the
Australian-American Educational Foundation.

\vfill\eject
\bigskip
\References

\bye